\documentclass[11pt]{article}
\usepackage{rotating}
\usepackage{epsfig}
\usepackage{latexsym}
\usepackage{graphicx}
\usepackage{amsmath,amssymb,slashed}
\def\ltsim{\lower3pt\hbox{$\, \buildrel < \over \sim \, $}}  
\def\gtsim{\lower3pt\hbox{$\, \buildrel > \over \sim \, $}}  
\makeatletter
\def\section{\@startsection {section}{1}{\z@}{-3.5ex plus -1ex minus
 -.2ex}{2.3ex plus .2ex}{\large\bf}}
\def\subsection{\@startsection{subsection}{2}{\z@}{-3.25ex plus -1ex
minus -.2ex}{1.5ex plus .2ex}{\normalsize\bf}}
\makeatother

\newcommand{\captionfonts}{\small}
\makeatletter  
\long\def\@makecaption#1#2{%
  \vskip\abovecaptionskip
  \sbox\@tempboxa{{\captionfonts #1: #2}}%
  \ifdim \wd\@tempboxa >\hsize
    {\captionfonts #1: #2\par}
  \else
    \hbox to\hsize{\hfil\box\@tempboxa\hfil}%
  \fi
  \vskip\belowcaptionskip}
\makeatother   

\catcode`@=11
\def\marginnote#1{}
\newcount\hour
\newcount\minute
\newtoks\amorpm
\hour=\time\divide\hour
by60
\minute=\time{\multiply\hour by60 \global\advance\minute
by-\hour}
\edef\standardtime{{\ifnum\hour<12 \global\amorpm={am}
\else\global\amorpm={pm}\advance\hour by-12 \fi
 \ifnum\hour=0
\hour=12 \fi
 \number\hour:\ifnum\minute<10
0\fi\number\minute\the\amorpm}}
\edef\militarytime{\number\hour:\ifnum\minute<10
0\fi\number\minute}
\def\draftlabel#1{{\@bsphack\if@filesw
{\let\thepage\relax
 \xdef\@gtempa{\write\@auxout{\string
\newlabel{#1}{{\@currentlabel}{\thepage}}}}}\@gtempa
 \if@nobreak
\ifvmode\nobreak\fi\fi\fi\@esphack}
\gdef\@eqnlabel{#1}}
\def\@eqnlabel{}
\def\@vacuum{}
\def\draftmarginnote#1{\marginpar{\raggedright\scriptsize\tt#1}}
\def\draft{\oddsidemargin
0.0truein
 \def\@oddfoot{\sl preliminary draft \hfil
\rm\thepage\hfil\sl\today\quad\militarytime}
 \let\@evenfoot\@oddfoot
\overfullrule 3pt
 \let\label=\draftlabel
\let\marginnote=\draftmarginnote
\def\@eqnnum{(\theequation)\rlap{\kern\marginparsep\tt\@eqnlabel}
\global\let\@eqnlabel\@vacuum}
}
\catcode`@=12

\def\XXint#1#2#3{{\setbox0=\hbox{$#1{#2#3}{\int}$}
     \vcenter{\hbox{$#2#3$}}\kern-.5\wd0}}

\def\bea{\begin{align}} 
\def\eea{\end{align}}
\def\be{\begin{equation}} \def\ee{\end{equation}} 

\newcommand{\promille}{%
  \relax\ifmmode\promillezeichen
        \else\leavevmode\(\mathsurround=0pt\promillezeichen\)\fi}
\newcommand{\promillezeichen}{%
  \kern-.05em%
  \raise.5ex\hbox{\the\scriptfont0 0}%
  \kern-.15em/\kern-.15em%
  \lower.25ex\hbox{\the\scriptfont0 00}}

\begin{document}

\thispagestyle{empty}

\begin{flushright}
\small{UCSD PTH/11-01}
\end{flushright}

\begin{center}
\vspace*{2cm}
\begin{center}

\vspace{1.7cm}

{\LARGE\bf Ultraviolet Properties of the Higgs Sector\vspace*{0.5cm}\\
in the Lee-Wick Standard Model}

\end{center}

\vspace{1.4cm}

{\large\bf Jos\'e R. Espinosa}$^a$ and {\large\bf Benjam\'{\i}n 
Grinstein}$^b$\\

\vspace{1.2cm}

$^a${\em {ICREA, Instituci\`o Catalana de Recerca i Estudis Avan\c{c}ats,
Barcelona, Spain}}\\

{\em {at  IFAE, Universitat Aut{\`o}noma de Barcelona,
08193 Bellaterra, Barcelona, Spain}}

$^b${\em Department of Physics, University of California, San Diego,
  CA 92093 USA}

\end{center}

\vspace{0.8cm}

\centerline{\bf Abstract}
\vspace{2 mm}
\begin{quote}\small
The Lee-Wick (LW) Standard Model (SM) offers a new solution to the 
hierarchy problem. We discuss, using effective potential techniques, 
its peculiar ultraviolet (UV) behaviour. We show how quadratic 
divergences in the Higgs mass $M_h$ cancel as a result of the unusual 
dependence of LW fields on the Higgs background (in a manner reminiscent 
of Little Higgses). We then extract from the effective potential the 
renormalization group evolution of the Higgs quartic coupling $\lambda$ 
above the LW scale. After clarifying an apparent discrepancy with previous
results for the LW Abelian Higgs model we focus on the LWSM.
In contrast with the SM case, for any $M_h$, $\lambda$ 
grows monotonically and hits a Landau pole at a fixed trans-Planckian 
scale (never turning negative in the UV). 
Then, the perturbativity and stability bounds on $M_h$ disappear.
We identify a cutoff $\sim 10^{16}$ GeV for the LWSM due to the hypercharge
gauge coupling hitting a Landau pole. Finally, we also discuss briefly the 
possible impact of the UV properties of 
the LW models on their behaviour at finite temperature, in particular 
regarding symmetry nonrestoration.
 \end{quote}

\vfill

\newpage

\section{Introduction}

In an effort to tame the divergences of quantum field theory Dirac
proposed a formulation of quantum mechanics with indefinite metric in
the Hilbert space \cite{dirac}. Pauli further studied Dirac's proposal
and found it to be effective in eliminating certain divergences but
failed to give a consistent interpretation of the theory
\cite{pauli}. Pauli and Villars showed that Lagrangians with
derivatives higher than of the second order are equivalent to negative
metric theories without higher derivatives \cite{paulivillars}. They
introduced the now famous regulator procedure in which the negative
metric states are rendered arbitrarily heavy at the end of the
computation.  Two decades later, motivated by their desire to
eliminate infinities in QED, Lee and Wick (LW) proposed a solution to
the question of interpretation of negative metric quantization
\cite{LW}. They argued that under certain conditions a theory of this
kind has a unitary $S$-matrix. Physically their proposal is that
states that in the absence of interactions are of negative metric may
well be unstable when interactions are present and sufficiently
strong. Since unstable states are not asymptotic,  only the subspace of
the Hilbert space corresponding to positive metric contributes to the
$S$-matrix.

The work of 't~Hooft and Veltman on renormalization of gauge theories
shelved the LW proposal for a decade, but it was dusted with growing
interest in quantizing gravity. In particular it was shown that a
higher derivative version of Einstein's theory of relativity is
renormalizable \cite{Stelle:1976gc} and asymptotically
free \cite{Fradkin:1981hx}. More recently it was realized that the
Higgs mass in higher derivative versions of the Standard Model (SM) of
electroweak interactions does not suffer from a quadratic divergence
\cite{GOW}. Instead there is only logarithmic sensitivity to the
cut-off, and the shift in the Higgs mass is of order $M^2/16\pi^2$,
where $M$ is the scale that characterizes the higher derivatives.  The
result remains valid even if the model is extended to incorporate
right handed neutrinos with masses much large than $M$, that generate
light Majorana masses via the see-saw mechanism \cite{neut}. 

This ``Lee-Wick Standard Model'' (LWSM) is consistent with electroweak
precision data \cite{pe} and with flavor physics constraints provided
$M$ is at least a few TeV \cite{dulaney}. The electroweak data favors
a light Higgs, $m_h\sim 100$ -- 200~GeV, which remarkably requires
little if any finetuning for $M$ a few TeV. Such low values for
$M$ have observable effects in collider experiments. At the LHC one
would expect to see resonances \cite{rizzo} associated with would-be
negative metric states, roughly one per SM particle. 

While this successful yet natural phenomenology is encouraging, there
remain many questions of principle with regard to the Lee-Wick
proposal. Whether the LW proposal yields a unitary theory is unknown
in general.  Cutkosky, Landshoff, Olive and Polkinghorne sharpened the
prescription of Lee and Wick and showed that large classes of diagrams
in perturbation theory satisfy the cutting relations needed for
perturbative unitarity \cite{CLOP}. Yet, for some specific models
unitarity can be shown to hold explicitly to all orders
\cite{tomboulis,Grinstein:2008bg}.  Boulware and Gross have identified
difficulties with a path integral formulation of the theory \cite{bg},
but van Tonder has recently proposed a non-perturbative definition for
the theory \cite{tonder}. And, already known to Lee and Wick, their
quantization procedure gives non-local correlations that are readily
interpreted as non-causal effects. 

These non-causality is readily seen as time advancement in certain
scattering processes. To be sure, in the LWSM, with the scale $M$ of
order of a few TeV,  these time advancements are unmeasurably short at
present. A question immediately arises as to whether a macroscopic
sequence of non-causal effects could be contrived to produce
macroscopic violations of cause and effect, rendering these theories
inconsistent. Coleman argued that this is not possible, but gave no
detailed argument \cite{coleman}. An attempt to address this question
indirectly was made in Ref.~\cite{Fornal:2009xc}, where the behavior
of LW models at high temperature was studied. The speed of sound was
found to increase with temperature but never to exceed the speed of
light. However, a somewhat surprising and discouraging effect was
discovered. The energy density of a LW gas of fermions was determined
to decrease without bound as the temperature is increased.  

Intending to throw some light into this problem we propose as a first step 
to investigate the effective potential of LW theories with scalars, 
fermions and gauge bosons. In Sec.~\ref{sec:LWVeff} we examine the UV 
behaviour of the Coleman-Weinberg effective potential in such generic LW 
theories, using for simplicity the higher-derivative formulation and 
Landau gauge (discussing in turn the contributions to the potential of 
generic bosonic and fermionic degrees of freedom). In order to show the 
cancellation of some UV divergences, we find convenient to regularize the 
potential using a momentum cutoff. Similarities between Lee-Wick and 
Little-Higgs theories show up most clearly in this language. We also 
investigate the finite part of the potential and ask, for example, under 
what conditions it may have runaway directions.

As a by-product, from the effective potential we are able to determine 
some renormalization group equations (RGEs) in specific models: making use 
of the renormalization-scale independence of the effective potential (and 
the knowledge of the scalar anomalous dimension) it is possible to extract 
from the one-loop effective potential the RGEs of the parameters of the 
tree level potential (mass terms and quartic coupling). RGEs for 
Yang-Mills LW models with fermions and scalars were determined in 
Ref.~\cite{LWren}. The models did not include scalar self-couplings and 
the calculations were performed in Landau gauge. The LW Abelian-Higgs 
model for arbitrary $\xi$-gauge, including a scalar self-coupling was 
computed in Ref.~\cite{Chivukula:2010kx} with the surprising result that 
the beta function of the scalar self-coupling vanishes. Our computation of 
the effective potential gives results at odds with 
Ref.~\cite{Chivukula:2010kx}. In particular, we find that the quartic 
self-coupling does run. To fully clarify and settle this issue, we present 
four independent calculations of the RGEs of the model, to wit, by 
computing green functions diagrammatically (Sec.~\ref{sec:RGE}) or by 
computing the effective potential (Sec.~\ref{sec:RGE-V}), and in both 
cases in the higher derivative and auxiliary-field formulations. To show 
explicitly that calculations in different formulations agree requires 
matching correctly the parameters of both formulations and dealing with a 
subtlety in the treatment of anomalous dimensions in the auxiliary-field 
formulation of the model. In the end, the discrepancy with 
\cite{Chivukula:2010kx} is only apparent and due to a different 
renormalization prescription.

Finally, Sec.~\ref{sec:disc} discusses the implications of the softer UV 
behaviour of the LW scalar sector in the context of the LW Standard Model. 
First, we derive in Sec.~\ref{sec:impl0} the RGEs of the 
parameters of the Higgs sector in the LWSM, with particular attention to 
the Higgs quartic coupling. We find that the running of this coupling is 
better UV-behaved than in the normal SM: it does not get driven to 
negative values at high energy if the Higgs mass is low nor does it 
blow-up below the Planck mass if the Higgs mass is large. As a result, in 
the LWSM the lower stability bound and the perturbativity bound on the 
Higgs mass disappear. Nevertheless, the RG evolution of gauge couplings 
above the LW mass scale $M$ is also modified \cite{LWren} and we find a 
Landau pole for the $U(1)_Y$ gauge coupling at a scale $\Lambda'\sim 
10^{16}$ GeV (for $M\sim 1$ TeV). This implies that the pure LWSM cannot 
be extrapolated up to the Planck scale and new physics should 
appear at or below $\Lambda'$.

At finite temperature (Sec.~\ref{sec:implT}) there is another 
reason why the ultraviolet behavior of the LW effective potential is of 
interest.  In Little Higgs models EW symmetry can remain broken at high 
temperature. More generally, symmetry non-restoration can occur in models 
for which quadratic divergences in the Higgs mass cancel among states with 
same statistics \cite{Espinosa:2004pn}. Heuristically, this is because 
$T^2m^2$ corrections to the finite temperature effective potential, 
which are responsible for symmetry restoration, are directly related to 
quadratic divergences to the Higgs mass at zero temperature. Since in LW 
theories cancellation of divergences are among states with same statistics 
we should then find that EW symmetry might not get restored at high 
temperature. However, it is not immediately obvious how to extend the 
standard calculation of the finite temperature effective potential to LW 
models. At any rate, the above argument indicates that the fate of 
symmetry at high temperature is determined by the sensitivity of the 
effective potential to the ultraviolet.

\section{Structure of the LW Effective Potential}
\label{sec:LWVeff}
In order to compute the effective potential in LW theories, we choose to 
do the computation using a simple momentum cutoff to regularize divergent 
integrals. This makes the UV behavior, in particular the absence of 
quadratic divergences, more readily apparent.

\subsection{UV Behaviour of the Effective Potential}
As preparation for the computation in LW theories, let us begin by
revisiting the normal (non-LW) case.  Consider a theory of
a single self-interacting scalar, 
\be
\mathcal{L}=\frac12(\partial_\mu \phi)^2 -V_0\ ,
\ee
with a SM-like Higgs sector with  tree-level potential
\be
\label{V0}
V_0=\frac{1}{2}\mu^2\phi^2+\frac{1}{4}\lambda \phi^4\,.
\ee
In the presence of a uniform background, $\phi(x)= v+h(x)$, the
one-loop vacuum diagrams are each infrared divergent. The divergence is,
however, an artifact of perturbation theory and the IR finite 
sum gives the effective potential:
\be
\label{Vsm}
V_1=\frac{1}{32\pi^2} \int_0^{\Lambda^2}p_E^2\ 
dp_E^2\ \log(p_E^2+m^2)\,.
\ee
Here $m^2\equiv d^2V_0/d\phi^2|_{\phi=v}$ is the mass in the
non-vanishing uniform background. The result is readily generalized 
to theories of many fields, including scalars, fermions and gauge bosons: 
\be
\label{Vsm-general}
V_1=\frac{1}{32\pi^2}\sum_\alpha N_\alpha \int_0^{\Lambda^2}p_E^2\ 
dp_E^2\ \log(p_E^2+m_\alpha^2)\,,
\ee
where the sum is over particle species $\alpha$ with $N_\alpha$ degrees 
of freedom (negative for fermions)
and mass $m_\alpha$ (dependent in general on the Higgs field
background) and $p_E$ is the Euclidian momentum. 

Although one could integrate (\ref{Vsm}) exactly, we can readily
extract the dominant UV behaviour simply by taking the derivative of
$V_1$ with respect to $m_\alpha^2$, doing the momentum integral and
then integrating in $m_\alpha^2$. In this way, one gets 
\be
\label{VsmUV}
\begin{split}
V_1 &= \frac{1}{32\pi^2}\sum_\alpha N_\alpha \left[\Lambda^2
  m_\alpha^2 -\frac{1}{2}m_\alpha^4\log\Lambda^2+...
\right]\\
&\equiv  \frac{1}{32\pi^2} \left[\Lambda^2\ {\mathrm{Str}}{\cal M}^2
  -\frac{1}{2}\ {\mathrm{Str}}{\cal M}^4\log\Lambda^2+\ldots  \right]\,,
\end{split}
\ee
where the dots stand for finite terms or terms suppressed by inverse
powers of the cutoff. We have used the super-trace, Str, to denote the
trace of a matrix weighted by the number of degrees of
freedom, and $\mathcal{M}$ stands for a matrix of masses of all fields
in the background of the Higgs fields.  As usual, the logarithmic
dependence on the cutoff tracks the RG evolution of the parameters of
the model. In section~\ref{sec:RGE-V} we will use effective potential
expressions like these to derive RGEs in LW models.

\subsection{Bosonic Contributions to the LW Effective Potential}
\label{sec:LWVeffB}

We now turn to the case of LW theories. Take for definiteness the case
of scalar fields. The Lagrangian is now 
\be
\mathcal{L}=\frac12(\partial_\mu \phi_\alpha)^2
-\frac1{2M^2}(\partial^2\phi_\alpha)^2 -V_0 \ .
\ee 
The Feynman rules now have quadratic polynomials in $p^2$ in
propagator denominators. Repeating the steps that lead to Eq.~\eqref{Vsm}
one finds instead:
\be
\label{Vblw}
V_1=\frac{1}{32\pi^2}\sum_\alpha N_\alpha \int_0^{\Lambda^2}p_E^2\
dp_E^2\ \log(p_E^2+m_\alpha^2+p_E^4/M^2)\,.  
\ee 
The Wick-rotation to Euclidean momentum is justified by the Lee-Wick
prescription for the contour of integration in the complex energy
plane. That is, first, in the theory with interactions switched off, take the
usual Feynman contour, just above or below the real axis as determined
by the $i\epsilon$ prescription. Then deform it to avoid crossing
the poles that migrate into the complex energy plane as the
interactions in the LW model are switched on.

This generic form is also applicable to gauge bosons in Landau gauge. 
In LW theory, for each gauge field, supplement the Lagrangian with a 
term $\frac1{2M^2}[(D^\mu F_{\mu\nu})^a]^2$. In a later section we 
discuss the more general case with a renormalizable gauge-fixing. 
The main points presented in this section are not affected by sticking 
to the simpler Landau gauge.

Just as above, the integral is most easily performed by differentiating and
integrating with respect to masses. One gets the following
UV behaviour 
\be
\label{VblwUV}
V_1=\frac{1}{32\pi^2}\sum_\alpha N_\alpha \left[ m_\alpha^2 
\ M^2\log\Lambda^2+...\right]\,.
\ee
Comparing with (\ref{VsmUV}), we immediately see striking dissimilarities:
there is no quadratic divergence and the structure of the logarithmic
divergence is quite different. The latter has the structure of the
quadratic divergence of the normal case if $M$ were the cutoff. This
is expected since in a scalar theory the higher derivatives could  be
used as a regulator. 

In order to better understand this result it is useful to look at the
LW theory in terms of new auxiliary LW degrees of freedom added to a standard
theory. In this formulation terms higher than quadratic in derivatives
are absent from the Lagrangian. Instead, these extra LW fields
are responsible for the additional poles of the modified propagator.
That is,  the masses of the normal and auxiliary LW fields
correspond to the solutions of the pole equation: 
\be
\label{poleb}
p^4-p^2\ M^2 +M^2\ m_\alpha^2=0\,, 
\ee 
and the LW field is identified by the pole with negative residue. This
corresponds to a wrong sign kinetic energy term in the
Lagrangian. Alternatively one can make the sign of the kinetic term of
the auxiliary LW field standard by rescaling the field by $i$. Then
the structure of these pole masses can also be obtained as coming from
a non-hermitian mass matrix of the form 
\be
\label{massb}
{\cal M}^2_{B \alpha}=\left[\begin{array}{cc}
m_\alpha^2 & -i\ m_\alpha^2 \\
-i\ m_\alpha^2 & M^2 -m_\alpha^2
\end{array}
\right]\,.	
\ee
The two solutions of the pole equation (\ref{poleb}), or, equivalently,
the two eingenvalues of the mass matrix (\ref{massb}), are
\be
\label{massesb}
M_{B \alpha 1,2}^2 = \frac{M^2}{2}\left(1\mp 
\sqrt{1-\frac{4m_\alpha^2}{M^2}}\right)\,.
\ee
These two masses are real if $m_\alpha^2<M^2/4$. This holds for the
usual choice of parameters in applications of LW theory to the
hierarchy problem since $m_\alpha$ are of electroweak size while $M$
is taken in the several TeV range. When calculating the one-loop effective
potential for values of the Higgs field background for which
$m_\alpha^2>M^2/4$, the two masses (\ref{massesb}) are complex 
conjugate pairs, their sum giving a real contribution to the potential
(see below). Expanding the masses (\ref{massesb}) in powers of $m_\alpha^2/M^2$
we find
\be
\label{massesbexp}
\begin{aligned}
M_{B \alpha 1}^2 & =  m_\alpha^2 + {\cal O}(m_\alpha^4/M^2)\,, 
\\
M_{B \alpha 2}^2 & =  M^2-m_\alpha^2-m_\alpha^4/M^2+{\cal 
O}(m_\alpha^6/M^4)\,.
\end{aligned}
\ee

\begin{figure}
\center{
\includegraphics[width=10cm,height=8cm]{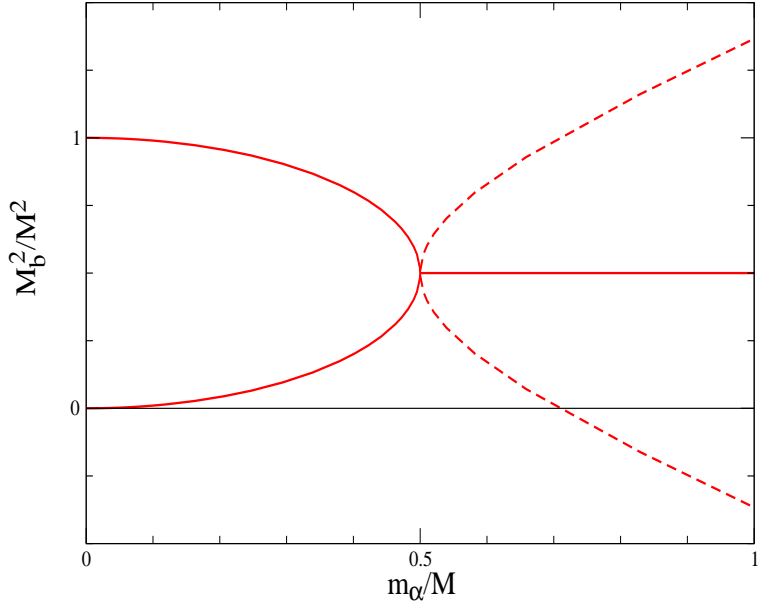}}
\caption{\label{fig:bosonmasses} Squared-masses of a bosonic
  ``LW-multiplet'' as a function of the ratio $m_\alpha/M$. See text
  for explanations.}
\end{figure}

Figure~\ref{fig:bosonmasses} shows the squared-masses for bosons
throughout both low and high Higgs background regions as a function of the
ratio $m_\alpha/M$. The two complex masses in the high region are
represented by plotting their real part, $M^2/2$, as a solid line while
the dashed lines give $M^2(1\pm\sqrt{4m_\alpha^2/M^2-1})/2$ as a
convenient way of plotting the information on the imaginary parts.

In summary, for each standard bosonic degree of freedom with mass
squared $m_\alpha^2$ (up to corrections suppressed by $M$) there is a
new LW degree of freedom with mass squared $M^2-m_\alpha^2+\ldots$
completing a ``LW-multiplet.'' Using the standard formula
(\ref{VsmUV}) for these degrees of freedom and keeping a unique label
$\alpha$ for each SM-LW pair we reproduce the UV behaviour of
(\ref{VblwUV}), up to an irrelevant background-field independent
constant. One sees explicitly that this is the result of cancellations
between the normal and LW contributions. We can also see this
cancellation as occurring through ${\mathrm{Tr}}[{\cal M}_{B
  \alpha}^2]$ and ${\mathrm{Tr}}[{\cal M}_{B \alpha}^4]$ directly
without any power expansion, which is one of the uses of
writing down the mass matrix ${\cal M}^2_{B\alpha}$ in (\ref{massb}).
One can also see the two contributions arising directly from 
 the integral (\ref{Vblw})  by factoring the argument of the logarithm:
\be
\label{eq:VFromtwoMBs}
V_1=\frac{1}{32\pi^2}\sum_\alpha N_\alpha \int_0^{\Lambda^2}p_E^2\ 
dp_E^2\ \log[(p_E^2+M_{B \alpha 1}^2)(p_E^2+M_{B \alpha 2}^2)]\,.
\ee
Explicitly, the contribution to the one-loop potential reads
\be
\delta_\alpha V_1=\frac{N_\alpha}{64\pi^2}\sum_{i=1,2}M_{B\alpha 
i}^4\left[\log\frac{M^2_{B\alpha i}}{Q^2}-C_\alpha\right]\,,
\ee
where we have merely used the standard Coleman-Weinberg expression in
Landau gauge. Here $C_\alpha=5/6$ for gauge bosons, $C_\alpha=3/2$ for
scalars and $Q$ is the renormalization scale.

In the low field region, for which $m_\alpha^2<M^2/4$, the potential above
takes the form
\be
\label{Vblow}
\delta_\alpha 
V_1=\frac{N_\alpha}{64\pi^2}M^4\left[\left(1-\frac{2m_\alpha^2}{M^2}\right)
\left(\log\frac{M m_{\alpha}}{Q^2}-C_\alpha\right)-\frac12\zeta_\alpha 
\log\frac{1-\zeta_\alpha}{1+\zeta_\alpha}\right]\, ,
\ee
with
\be
\zeta_\alpha\equiv \sqrt{1-\frac{4m_\alpha^2}{M^2}}\,.
\ee

By contrast, in the high field region, for which $m_\alpha^2>M^2/4$,
the potential above takes the form\footnote{Note that 
         (\ref{Vblow}) is simply 
         the analytic continuation of (\ref{Vbhigh})
         into $\zeta_\alpha=i\Delta_\alpha<1$.}
\be
\label{Vbhigh}
\delta_\alpha 
V_1=\frac{N_\alpha}{64\pi^2}M^4\left[\left(1-\frac{2m_\alpha^2}{M^2}\right)
\left(\log\frac{M m_{\alpha}}{Q^2}-C_\alpha\right)-\Delta_\alpha\ 
\arctan\Delta_\alpha\right]\,,
\ee
with
\be
\Delta_\alpha\equiv \sqrt{\frac{4m_\alpha^2}{M^2}-1}\ .
\ee

We can see from this result that the bosonic contributions to the one-loop
effective potential now grow only like $v^2\log(v^2)$ for high $v\gg M$
(to be compared with the $v^4$ growth in the normal case).  Therefore, in that
region of field space the tree level term $(\lambda/4) v^4$ will dominate.

\subsection{Fermionic Contributions to the LW Effective Potential}
\label{sec:LWVeffF}
We consider next the contributions of fermions to the effective potential 
in a theory with higher derivatives. The terms in the
Lagrangian with derivatives on fermions are
\be
\mathcal{L}=\bar\psi_\alpha i\slashed{\partial}
(1-\partial^2/M^2)\psi\,. 
\ee
The fermionic contributions to the effective potential in LW theories
can be obtained through an analysis similar to the one for bosons
above. Instead of (\ref{Vsm}), for fermions in LW theories one has
\be
\label{Vflw}
V_1=\frac{1}{32\pi^2}\sum_\alpha N_\alpha \int_0^{\Lambda^2}p_E^2\ 
dp_E^2\ \log\left[p_E^2(1+p_E^2/M^2)^2 + m_\alpha^2\right]\,,
\ee
where, we remind the reader, we have included a minus sign in
$N_\alpha$.
By  applying the same procedure as above one finds that fermions do not
contribute to the potential a field-dependent UV divergence, not even
logarithmic.

Again, we can understand this result in terms of new auxiliary LW
fermionic degrees of freedom. Now, the equation giving the propagator
poles reads
\be
\label{polef}
p^2(p^2-M^2)^2-m_\alpha^2\ M^4 = 0\,,
\ee
where $m_\alpha$ is the mass of the standard fermionic degree of freedom.
This pole equation admits now two additional solutions, corresponding to two
additional LW degrees of freedom.  The structure of these pole masses can
be obtained in an equivalent manner as coming from a non-hermitian
mass-squared matrix of the form\footnote{We are assuming that both LW
fields appear with the same heavy mass $M$ for simplicity, but this is not
necessary. In the most general case the LW mass in the 33 entry in 
(\ref{massf}) can be different from the other LW mass.}
\be
\label{massf}
{\cal M}^2_{F \alpha}=\left[\begin{array}{ccc}
M^2 & -i\ m_\alpha\ M & m_\alpha\ M \\
-i\ m_\alpha\ M & 0 & 0 \\ 
m_\alpha\ M & 0 & M^2 
\end{array}
\right]\,,	
\ee
which is what one would obtain by rescaling the auxiliary LW fields $\psi$ by
$\psi\to i\psi$ and $\bar\psi\to i\bar\psi$, which gives a standard 
sign for their kinetic term.

The eigenvalues of this matrix, or the solutions to the pole
equation, Eq.~(\ref{polef}), are
\be
\begin{aligned}
M_{F \alpha 1}^2 & =  \frac{M^2}{3} 
\left[\frac{}{}2-2\cos(\theta_\alpha/3)\right]\,,\nonumber\\
M_{F \alpha 2}^2 & =  
\frac{M^2}{3}\left[\frac{}{}2+\cos(\theta_\alpha/3)-
\sqrt{3}\sin(\theta_\alpha/3)\right]\,,\nonumber\\
M_{F \alpha 3}^2 & =  
\frac{M^2}{3}\left[\frac{}{}2+\cos(\theta_\alpha/3)+
\sqrt{3}\sin(\theta_\alpha/3)\right]\,,
\end{aligned}
\ee
where the angle $\theta_\alpha$ is given by
\be
\cos\theta_\alpha =1-\frac{27}{2}\frac{m_\alpha^2}{M^2}\,.
\ee
We assume here that $m_\alpha^2<4M^2/27$, which guarantees real masses.  
An expansion in powers of $m_\alpha^2/M^2$ gives
\be
\label{massesf}
\begin{aligned}
M_{F \alpha 1}^2 & =  m_\alpha^2+{\cal O}(m_\alpha^4/M^2)\,,\\
M_{F \alpha 2}^2 & =  M^2 - M m_\alpha - \frac{1}{2}m_\alpha^2 - 
\frac{5m_\alpha^3}{8M}-\frac{m_\alpha^4}{M^2}
+{\cal O}(m_\alpha^6/M^4)\,,\\
M_{F \alpha 3}^2 & =  M^2 + M m_\alpha - \frac{1}{2}m_\alpha^2 + 
\frac{5m_\alpha^3}{8M}-\frac{m_\alpha^4}{M^2}+{\cal O}(m_\alpha^6/M^4)\,.
\end{aligned}
\ee
Therefore, each standard fermionic degree of freedom is accompanied by two
quasidegenerate heavy LW-fields completing a fermionic ``LW-multiplet.''

Using the standard formula (\ref{VsmUV}) for the contribution of these
degrees of freedom to the one-loop potential and keeping a unique label
$\alpha$ for each standard-LW fermionic multiplet, we reproduce (up to a
field-independent constant) the UV finiteness of Eq.~(\ref{Vflw}) as a result
of standard-LW cancellations. The same cancellations can be seen as operating
directly in ${\mathrm{Tr}}[{\cal M}_\alpha^2]$ and ${\mathrm{Tr}}[{\cal
M}_\alpha^4]$ for the fermionic mass matrix (\ref{massf}). At the level of 
the integral (\ref{Vflw}) the three separate contributions to the
effective potential follow simply from  writing
the argument of the logarithm in factorized form:
\be
V_1=\frac{1}{32\pi^2}\sum_\alpha N_\alpha \int_0^{\Lambda^2}p_E^2\
dp_E^2\ \log[\Pi_{i=1,\ldots,3}(p_E^2+M_{F \alpha i}^2)]\,.
\ee

The explicit expression for the potential is
\be
\label{v1f}
\delta_\alpha V_1=\frac{N_\alpha}{64\pi^2}\sum_{i=1,2,3}M_{\alpha 
i}^4\left[\log\frac{M^2_{\alpha i}}{Q^2}-C_\alpha\right]\,,
\ee
where now $C_\alpha=3/2$. In fact, the only dependence
on $Q$ that appears in (\ref{v1f}) affects the renormalization of a 
background-field
independent term. For the purpose of studying the shape of the
background-field dependent potential, we can therefore simply drop $Q$
and $C_\alpha$  altogether in that expression.

In the high-field region, for which $m_\alpha^2> 4 M^2/27$, one of the three
mass eigenvalues is still real while the other two form a  complex
conjugate pair. They are
\be
\begin{aligned}
\label{MFs}
M_{F \alpha 1}^2 & = 
\frac{M^2}{2}\left\{\frac{4}{3}-f_+\left(\frac{m_\alpha}{M}\right)-
f_-\left(\frac{m_\alpha}{M}\right)
+i\sqrt{3}\left[f_+\left(\frac{m_\alpha}{M}\right)
-f_-\left(\frac{m_\alpha}{M}\right)\right]\right\}\,,\\
M_{F \alpha 2}^2 & =  \frac{M^2}{2}\left\{\frac{4}{3}
-f_+\left(\frac{m_\alpha}{M}\right)-
f_-\left(\frac{m_\alpha}{M}\right)
-i\sqrt{3}\left[f_+\left(\frac{m_\alpha}{M}\right)
-f_-\left(\frac{m_\alpha}{M}\right)\right]\right\}\,,\\
M_{F \alpha 3}^2 & =  
M^2\left[\frac{2}{3}+f_+\left(\frac{m_\alpha}{M}\right)+
f_-\left(\frac{m_\alpha}{M}\right)\right]\,,
\end{aligned}
\ee
where we have used the functions 
\be
f_{\pm}(x)\equiv \sqrt[3]{\frac{x^2}{2}
-\frac{1}{27}\pm x\sqrt{\frac{x^2}{4}-\frac{1}{27}}}\,.
\ee
For later use, we quote the useful relation $f_+(x)f_-(x)=1/9$. 

\begin{figure}
\center{
\includegraphics[width=10cm,height=8cm]{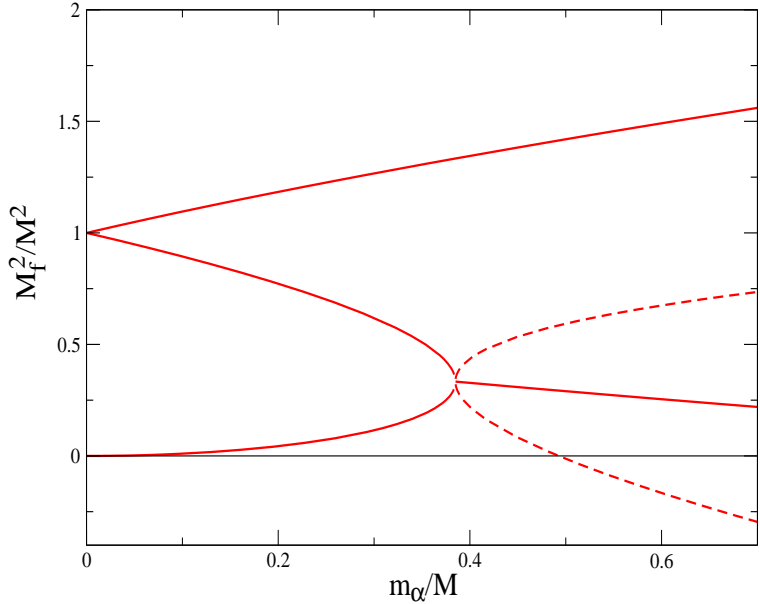}}
\caption{\label{fig:fermionmasses} 
Squared-masses of a fermionic ``LW-multiplet'' as a function of the ratio 
$m_\alpha/M$. The complex masses in the high region are represented
by plotting $M_{F\alpha }^2$ as a solid line and $M_{F\alpha }^2\pm
\Delta_{F\alpha}^2$ as dashed lines; see Eqs.~\eqref{MFs} and \eqref{v1fl}.}
\end{figure}

In the high field region, $m_\alpha^2> 4 M^2/27$,  the effective
potential takes the form: 
\be
\label{v1fl}
\delta_\alpha V_1  =  \frac{N_\alpha}{64\pi^2}\left[
M_{F\alpha 3}^4\log(M^2_{F\alpha 
3})+2(M_{F\alpha}^4-\Delta_{F\alpha}^4)\log(\rho_{F\alpha}^2)
-4M_{F\alpha}^2\Delta_{F\alpha}^2\theta_{F\alpha} \right]\,,
\ee
where 
\be
\begin{aligned}
M_{F\alpha}^2 & \equiv  
\frac{M^2}{2}\left[\frac{4}{3}-f_+\left(\frac{m_\alpha}{M}\right)-
f_-\left(\frac{m_\alpha}{M}\right)\right]\,,\\
\Delta_{F\alpha}^2 & \equiv  
\frac{M^2}{2}\sqrt{3}\left[f_+\left(\frac{m_\alpha}{M}\right)-
f_-\left(\frac{m_\alpha}{M}\right)\right]\,,\\
\rho_{F\alpha}^4 & \equiv  M_{F\alpha}^4 + \Delta_{F\alpha}^4\,,\\
\theta_{F\alpha} & \equiv  
\arctan{\frac{\Delta_{F\alpha}^2}{M_{F_\alpha}^2}}\,,
\end{aligned}
\ee
that is, $M_{F \alpha 1,2}^2=M_{F\alpha }^2 \pm i 
\Delta_{F\alpha}^2=\rho_{F\alpha}^2 \exp{(i\theta_{F\alpha})}$.
Different fermionic contributions in Eq.~(\ref{v1fl}) grow at high $v\gg M$ as
$v^{4/3}\log(v^2)$ and $v^{4/3}$. There is a cancellation of the dominant 
$v^{4/3}\log(v^2)$ terms, leaving a total result that grows only as 
$v^{4/3}$.
These contributions are therefore subdominant compared with the
tree-level quartic.

Figure~\ref{fig:fermionmasses} shows the squared-masses for fermions
throughout both low and high  background field regions as a function of the
ratio $m_\alpha/M$. The complex masses in the high region are represented
by plotting $M_{F\alpha }^2$ as a solid line and $M_{F\alpha }^2\pm
\Delta_{F\alpha}^2$ as dashed lines.

\section{RGEs in the LW Abelian Higgs Model. Diagrammatic Approach}
\label{sec:RGE}
As a warm-up for the LWSM case, in this section we calculate the 
renormalization group equations
(RGEs) of the scalar sector parameters in the Lee-Wick Abelian Higgs
model. We do this by computing directly the one-loop counterterms needed 
to
renormalize Green functions. We compute them first, in Sec.~\ref{sec:diagHD},
using the higher-derivative formulation of the model and then we
calculate them again, in Sec.~\ref{diagAuxF}, using the auxiliary-field
formulation. We find agreement between both approaches, once the
parameters in the two formulations are appropriately matched to each
other. These results will be used as the benchmark against which the
effective potential calculation of the RGEs (Sec.~\ref{sec:RGE-V}) can be
compared. This model already captures the main features and
subtleties of the LWSM calculation (which we present in 
section~\ref{sec:disc}) with the advantage of being simpler.

\subsection{Diagrammatic Approach in the Higher-Derivative Formulation}
\label{sec:diagHD}
The Lagrangian of the LW Abelian Higgs model in the higher-derivative 
formulation (indicated
by hatted fields and parameters) reads:
\be
\label{Lhd}
{\cal L_{HD}}= -\frac{1}{4}\hat{F}_{\mu\nu}^2 + 
\frac{1}{2\hat{M}_A^2}(\partial^\mu \hat{F}_{\mu\nu})^2
-\frac{1}{2\xi}(\partial^\mu \hat{A}_\mu)^2
+ |\hat{D}_\mu \hat\phi|^2 - \frac{1}{\hat{M}^2} |\hat{D}^2 \hat\phi|^2 - 
\hat{m}^2 |\hat\phi|^2
-\hat{\lambda}|\hat\phi|^4\,,
\ee
where $\hat{D}_\mu \hat\phi\equiv \partial_\mu \hat\phi+
i g \hat{A}_\mu \hat\phi$ and we show explicitly  the gauge-fixing term. 
With this gauge-fixing
the gauge-boson propagator is
\be
P_{\mu\nu}(p)=\frac{-\hat{M}_A^2}{p^2(p^2-\hat{M}_A^2)}
\left[g_{\mu\nu}-\frac{p_\mu p_\nu}{p^2}\right]
+\xi \frac{p_\mu p_\nu}{p^4}\,.
\ee
The scalar propagator can be written as
\be
\label{scap}
P(p )=\frac{1}{\hat{m}^2-p^2+p^4/\hat{M}^2}
=\frac{\hat{M}^2}{(p^2-m_1^2)(p^2-m_2^2)}\,,
\ee
with $m_1^2+m_2^2=\hat{M}^2$ and $m_1^2 m_2^2 =\hat{m}^2\hat{M}^2$.

We are interested in calculating the RGEs of the parameters in the
scalar sector of the theory, that is, the beta functions of
$\hat{m}^2$, $\hat{M}^2$, $\hat\lambda$ and the anomalous dimension of
$\hat\phi$. A straightforward one-loop diagrammatic calculation using
dimensional regularization gives the following result for the
divergent piece of the scalar two-point function:
\be
\label{twopf}
16\pi^2\Pi(p )^{UV}=g^2 C_{UV}\left[-\xi \frac{p^4}{\hat{M}^2}
+\left(6\frac{\hat{M}_A^2}{\hat{M}^2}+\xi\right)p^2
-3\hat{M}_A^2-\xi\hat{m}^2\right]-4\hat\lambda C_{UV}\hat{M}^2\,,
\ee
where
\be
C_{UV}\equiv\frac{1}{\epsilon}-\gamma_E+\log(4\pi)\,,
\ee
with $\epsilon=(4-d)/2$ and $\gamma_E$ the Euler constant. From
Eq.~(\ref{twopf}) we can extract in the standard way the following
RGEs:
\begin{align}
\label{gammahat}
\gamma_{\hat\phi}&\equiv \frac{d\, \hat\phi}{d\, \log Q} = 
-\frac{g^2}{16\pi^2}\left(6\frac{\hat{M}_A^2}{\hat{M}^2}+\xi\right)\,,\\
\beta_{\hat{M}^2}&\equiv  \frac{d\, \hat{M}^2}{d\, \log Q} = 
-\frac{g^2}{16\pi^2} 12\hat{M}_A^2\,,
\label{betaMhat}\\
\beta_{\hat{m}^2}&\equiv \frac{d\, \hat{m}^2}{d\, \log Q} =
-\frac{6g^2}{16\pi^2}
\hat{M}_A^2\left(1-2\frac{\hat{m}^2}{\hat{M}^2}\right)
-\frac{8\hat\lambda}{16\pi^2}\hat{M}^2\,.
\label{betamhat}
\end{align}
At $\xi=0$ and $\hat\lambda=0$ these results are in accord with 
Ref.~\cite{LWren}.

In order to get the RGE for the scalar quartic coupling $\hat\lambda$
we need the divergent part of the four-point scalar function.  In the
limit of vanishing external momenta tending to zero, it reads
\be
\label{fourpf}
16\pi^2 \hat{L}^{UV}_0 = -2\xi\hat\lambda g^2 C_{UV}\ ,
\ee
where $\hat{L}_0$ is normalized as $\hat\lambda$.
Note that there are no contributions of order $\hat\lambda^2$ (the
corresponding diagrams are finite) or $g^4$ (UV divergences of
separate diagrams cancel out). From Eq.~(\ref{fourpf}) and the
previous result on the scalar anomalous dimension,
Eq.~(\ref{gammahat}), we  obtain 
\be
\beta_{\hat\lambda}\equiv  \frac{d\, \hat\lambda}{d\, \log Q} = 24 
\frac{g^2\hat\lambda}{16\pi^2}\, \frac{\hat{M}_A^2}{\hat{M}^2}\,.
\ee
This completes our task. As expected on general grounds \cite{GrossW}, 
the one-loop
beta functions for $\hat{m}^2$, $\hat{M}^2$ and $\hat\lambda$ are
gauge independent and only the scalar anomalous dimension depends on
the gauge-fixing parameter $\xi$.

\subsection{Diagrammatic Approach in the Auxiliary-Field Formulation}
\label{diagAuxF}
We now turn to the calculation of the RGEs in the auxiliary-field
formulation, with derivatives at most of second order. We need an
auxiliary-field Lagrangian equivalent to the Higher derivative one in
Eq.~\eqref{Lhd}, which we can get by adding
auxiliary fields through 
\be
{\cal L} = {\cal L_{HD}} - \frac{1}{2}\hat{M}_A^2\left(\tilde{A}_\nu - 
\frac{1}{\hat{M}_A^2}\partial^\mu \hat{F}_{\mu\nu}\right)^2+
\hat{M}^2\left|\tilde{\phi}'-\frac{1}{\hat{M}^2}\hat{D}^2\hat\phi\right|^2\,,
\ee
where ${\cal L_{HD}}$ is the higher-derivative Lagrangian in
Eq.~(\ref{Lhd}). Replacing  the field $\hat\phi$ through the change of 
variables
$\hat\phi=\phi'-\tilde\phi'$ and performing a symplectic rotation
\be
\left(\begin{array}{c}
\phi'\\
\tilde{\phi}'
\end{array}
\right)
=\left(
\begin{array}{cc}
\cosh \theta & \sinh \theta\\
\sinh \theta & \cosh \theta
\end{array}
\right)\
\left(\begin{array}{c}
\phi\\
\tilde{\phi}
\end{array}
\right)\,,
\ee
with
\be
e^{4\theta}=1-4\frac{\hat{m}^2}{\hat{M}^2}\,,
\ee
we obtain
\begin{multline}
\label{Laux}
{\cal L} = -\frac{1}{4}F_{\mu\nu}^2 + \frac{1}{4}\tilde{F}_{\mu\nu}^2 - 
\frac{1}{2}M_A^2\tilde{A}_{\mu}\tilde{A}^{\mu}
-\frac{1}{2\xi}(\partial^\mu A_\mu-\partial^\mu \tilde{A}_\mu)^2\\
+ |D_\mu \phi|^2 - |D_\mu \tilde\phi|^2 +M^2 |\tilde\phi|^2 - m^2 |\phi|^2
-\lambda|\phi-\tilde\phi|^4 \\
+g^2 \tilde{A}_{\mu}\tilde{A}^{\mu} (|\phi|^2-|\tilde\phi|^2) +
i g \tilde{A}_\mu \left[\tilde\phi (D^\mu \tilde\phi)^* - \phi (D^\mu 
\phi)^* - {\mathrm h.c.}\right]\,,
\end{multline}
where now $D_\mu=\partial_\mu +igA_\mu$.

The dictionary between the new parameters $M_A^2$, $M^2$, $m^2$ and
$\lambda$ appearing in Eq.~(\ref{Laux}) and the original 
parameters in ${\cal L_{HD}}$ is the following: 
\be
\begin{aligned}
\label{dict}
m^2 & =  \frac{1}{2} \hat{M}^2\left[1-\sqrt{1-4\hat{m}^2/\hat{M}^2}\right]\,,\\
M^2 & =  \frac{1}{2} \hat{M}^2\left[1+\sqrt{1-4\hat{m}^2/\hat{M}^2}\right]\,,\\
\lambda & = \frac{\hat\lambda}{1-4\hat{m}^2/\hat{M}^2}\,,
\end{aligned}
\ee
and the trivial equality $M_A^2 =  \hat{M}_A^2$. The inverse relations are:
\be
\begin{aligned}
\label{invdict}
\hat{M}^2 & =  M^2 + m^2 \,,\\
\hat{m}^2 & =  \frac{m^2 M^2}{M^2 + m^2}\,,\\
\hat\lambda & = \lambda\; \frac{(M^2 - m^2)^2}{(M^2 + m^2)^2}\,.
\end{aligned}
\ee
Note that $m^2$ and $M^2$ correspond to the pole masses $m_1^2$ and
$m_2^2$ of the higher-derivative scalar propagator, as given by
Eq.~(\ref{scap}).

Before we compute directly the RGE for the parameters of this model
($M^2$, $m^2$ and $\lambda$)  we
can obtain them indirectly by differentiating relations (\ref{dict})
and using the  corresponding RGEs for
the hatted parameters, calculated in the preceding subsection, and
then use the relations (\ref{invdict}) to express the results in
terms of unhatted parameters. In this way one arrives at
\begin{align}
\label{betaM}
\beta_{M^2}&\equiv  \frac{d\, M^2}{d\, \log Q} = 
-\frac{1}{16\pi^2}\left[6 g^2 M_A^2-8\lambda (M^2-m^2)\right]\,,\\
\label{betam}
\beta_{m^2}&\equiv  \frac{d\, m^2}{d\, \log Q} = 
-\frac{1}{16\pi^2}\left[6 g^2 M_A^2+8\lambda (M^2-m^2)\right]\,,\\
\label{betal}
\beta_{\lambda}&\equiv  \frac{d\, \lambda}{d\, \log Q} = 
-\frac{1}{16\pi^2}32\lambda^2\,.
\end{align}

We now proceed to verify that these results follow from direct
diagrammatic calculation in the auxiliary-field formulation.
Explicitly, the divergent part of the two-point functions are: 
\be
\begin{aligned}
16\pi^2\Pi(p )^{UV}_{\phi\phi}&=-g^2 C_{UV}\left[3 M_A^2+\xi (m^2-p^2)\right]
-4\lambda C_{UV}(M^2-m^2)\,,\\
16\pi^2\Pi(p )^{UV}_{\tilde\phi\tilde\phi}&=g^2 C_{UV}\left[3 M_A^2+\xi 
(M^2-p^2)\right]
-4\lambda C_{UV}(M^2-m^2)\,,\label{twopfaux}\\
16\pi^2\Pi(p )^{UV}_{\phi\tilde\phi}&=4\lambda C_{UV}(M^2-m^2)\,.
\end{aligned}
\ee
These divergences can be compensated by counterterms in the usual way. 
Although the renormalization of the kinetic terms is invariant under an 
$SO(1,1)$ rotation among the fields $\phi$ and $\tilde\phi$, as explained 
in \cite{Chivukula:2010kx}, such rotation introduces mixed mass terms. For 
this reason we can absorb the non-zero $\Pi(p )^{UV}_{\phi\tilde\phi}$, 
which requires a mixed $\phi$-$\tilde\phi$ counterterm, through an 
off-diagonal anomalous dimension (even if the divergence is 
momentum-independent).\footnote{\label{foot}Alternatively, one could 
introduce a new mass term in the potential, 
$\mu^2(\phi^*\tilde\phi+\tilde\phi^*\phi)$, but this can always be rotated 
away by a field redefinition. Our prescription can be reinterpreted in 
terms of a renormalization of the mixing angle $\theta$.} More explicitly, 
we obtain
\be
\label{gammaux}
 \frac{d}{d\, \log Q} 
\left(\begin{array}{c}
\phi \\
\tilde\phi
\end{array}\right)
 \equiv 
\left(\begin{array}{cc}
\gamma_{\phi \phi} & \gamma_{\phi\tilde\phi}\\
\gamma_{\tilde\phi\phi} & \gamma_{\tilde\phi\tilde\phi}
\end{array}\right)\left(\begin{array}{c}
\phi \\
\tilde\phi
\end{array}\right)=
-\frac{1}{16\pi^2}
\left(\begin{array}{cc}
\xi g^2 & 8 \lambda\\
8 \lambda & \xi g^2
\end{array}\right)\left(\begin{array}{c}
\phi \\
\tilde\phi
\end{array}\right)\ .
\ee
These anomalous dimensions reproduce $d\, \hat\phi/d\, \log Q$ of
Eq.~(\ref{gammahat}), as can be easily checked simply writing
$\hat\phi$ in terms of $\phi$ and $\tilde\phi$. With the use of these
anomalous dimensions we can also obtain the RGEs for $M^2$ and $m^2$
from Eqs.~(\ref{twopfaux}) obtaining precisely the results anticipated
by Eqs.~(\ref{betaM}) and (\ref{betam}).

In order to get the one-loop RGE for $\lambda$ it is enough to compute
the divergent part of the one-loop four-point function for $\phi$.  In
the limit of vanishing external momentum we obtain
\be
\label{fourpfaux}
16\pi^2 L^{UV}_0 = -2\xi\lambda g^2 C_{UV}\ ,
\ee
where $L_0$ is normalized as $\lambda$.
The divergent pieces of mixed $\phi$-$\tilde\phi$ four-point functions
are such that $|\phi-\tilde\phi|^4$ is the divergent operator in the
one-loop effective action, so that a single counterterm for $\lambda$
can absorb that divergence. Making use of the scalar anomalous
dimensions as given by Eqs.~(\ref{gammaux}) one obtains a beta
function for $\lambda$ that reproduces the result given in
Eq.~(\ref{betal}).

In Ref.~\cite{Chivukula:2010kx} a different result is found, namely 
$\beta_\lambda=0$. This is the result of renormalizing differently the scalar 
mass terms and wave-functions, along the lines of footnote~\ref{foot}. Using 
such prescription implies in particular that the Higgs quartic coupling in 
\cite{Chivukula:2010kx} differs from ours by an overall factor (that depends
on the field-mixing radiatively induced) and therefore runs differently.
While the prescription in \cite{Chivukula:2010kx} is simpler in the sense of 
having a non-running $\lambda$, it requires the introduction of an additional 
mass parameter, which is absent in our prescription. Needless to say, all 
physical predictions of the theory should be prescription-independent.

\section{RGEs in the LW Abelian Higgs Model. Effective Potential
  Approach} 
\label{sec:RGE-V}
In this section we will rederive the RGEs for the parameters of the
scalar sector in the LW Abelian Higgs Model via the Coleman-Weinberg
potential and the scalar anomalous dimensions of the scalar
field(s). The technique, based on the scale-independence of the
effective potential, is well known \cite{Jones}. Consider a model with 
SM-like tree-level potential
\be
V_0=\frac{1}{2}\mu^2h^2+\frac{1}{4}\lambda h^4\,.
\ee
The one-loop Coleman-Weinberg correction is
\be
\label{VCW}
V_1 = \frac{1}{64\pi^2}\sum_\alpha N_\alpha M_\alpha^4(h) 
\left[\log\frac{M_\alpha^2(h)}{Q^2}-C_\alpha\right]\,,
\ee
where the sum runs over species $\alpha$ with $h$-dependent
mass-squared $M_\alpha^2(h)$ and $N_\alpha$ degrees of freedom (taken
negative for fermions); $Q$ is the renormalization scale and
$C_\alpha=5/6\, (3/2)$ for gauge bosons (scalars or
fermions). Imposing one-loop RG invariance of $V_0+V_1$ one obtains the
relations
\be
\label{betamfromV}
\beta_{\mu^2}+2\gamma \mu^2 =
 \frac{1}{16\pi^2}
\left.\left[
\frac{\partial}{\partial h^2}
{\mathrm{Str}}{\cal M}^4
\right]\right|_{h=0}\equiv
 \frac{1}{16\pi^2}\sum_\alpha N_\alpha 
\left.\frac{\partial M_\alpha^4}{\partial h^2}\right|_{h=0}\,,
\ee
\be
\label{betalfromV}
\beta_{\lambda}+4\gamma \lambda  =
 \frac{1}{16\pi^2}
\left.\left[
\frac{\partial^2}{(\partial h^2)^2}
{\mathrm{Str}}{\cal M}^4
\right]\right|_{h=0}\equiv
 \frac{1}{16\pi^2}\sum_\alpha N_\alpha 
\left.\frac{\partial^2 M_\alpha^4}{(\partial h^2)^2}\right|_{h=0}\,,
\ee
where $\beta_x\equiv d x/d\log Q$ and $\gamma \equiv d \log h/d\log Q$, as 
usual.
For masses of the generic form $M_\alpha^2= \mu_\alpha^2+\kappa_\alpha 
h^2$ one then obtains
\be
\beta_{\mu^2}+2\gamma \mu^2  = \frac{1}{8\pi^2}\sum_\alpha N_\alpha 
\kappa_\alpha\mu_\alpha^2\,,
\label{RGsm1}
\ee
\be
\beta_{\lambda}+4\gamma \lambda  = \frac{1}{8\pi^2}\sum_\alpha N_\alpha 
\kappa_\alpha^2\,.
\label{RGsm2}
\ee 
This procedure can be generalized trivially to cases with mass 
mixing and/or several scalar fields.

In order to determine the beta functions it is necessary to calculate
the anomalous dimension(s) separately. For the case of the Abelian
Higgs model we will take them from the previous section. (In
Sec.~\ref{betaVAuxF} we will discuss the subtleties that arise due to mixing
of the anomalous dimensions of normal and LW scalars in the auxiliary
field formalism.)

For the purpose of calculating these beta functions in a given model
we do not need to calculate explicitly the $M_\alpha$'s because the
scale dependence of $V_1$ only involves ${\mathrm{Str}}{\cal M}^4$, see
(\ref{betamfromV}) and (\ref{betalfromV}). In
general, the $M_\alpha$'s in each sector of the theory are solutions,
$p^2=M_\alpha^2$, of polynomial secular equations of the general form:
\be
(p^2)^n + (p^2)^{n-1} a_1 + (p^2)^{n-2} a_2 + ... a_n = 0\,,
\ee
where the $a_i$ are functions of the background field $h$. Writing
formally this equation  as
\be
\Pi_{\alpha=1}^{n}(p^2-M_\alpha^2)=0\,,
\ee
we immediately get
\be
\label{recipe1}
{\mathrm{Tr}}{\cal M}^2 \equiv \sum_\alpha M_\alpha^2 = -a_1\,,\;\;\;
{\mathrm{Tr}}{\cal M}^4 \equiv \sum_\alpha M_\alpha^4 = a_1^2-2a_2\,.
\ee
We will use these equations in what follows, applying them sector by
sector, to compute the separate contributions to the supertrace
${\mathrm{Str}}{\cal M}^4$. 

Before embarking into that detailed calculation for the Abelian Higgs 
Model, we can apply this technique to a general LW theory in the simple 
Landau gauge and assuming a unique LW mass $M$ (the case considered in 
our previous analysis of LW effective potential contributions). Bosonic LW 
multiplets, with pole equation as in (\ref{poleb}), will contribute to  
${\mathrm{Str}}{\cal M}^4$ the piece
\be
\left(\delta_\alpha {\mathrm{Str}}{\cal M}^4 \right)_B = M_{B\alpha 1}^4+
M_{B\alpha 2}^4 = M^4 - 2 m_\alpha^2 M^2 \ ,
\ee 
while fermionic LW multiplets, with pole equation as in (\ref{polef}), 
will 
give the $h$-independent piece
\be
\left(\delta_\alpha {\mathrm{Str}}{\cal M}^4 \right)_F =
-\sum_{i=1}^3 M_{F\alpha i}^4=-2 M^4 \ .
\ee 
If we input these results in the general formulas (\ref{betamfromV}) and 
(\ref{betalfromV}) and use $m_\alpha^2=\mu_\alpha^2 + \kappa_\alpha h^2$ 
we get, instead of the standard RGEs given in
Eqs.~\eqref{RGsm1}--\eqref{RGsm2}, 
\begin{gather}
\beta_{\mu^2}+2\gamma \mu^2 = -\frac{1}{8\pi^2}M^2{\sum_\alpha}' N_\alpha 
\kappa_\alpha\,,
\label{RGlw1}\\
\beta_{\lambda}+4\gamma \lambda  = 0\,,
\label{RGlw2}
\end{gather}
where the primed sum indicates that only bosons contribute and $\alpha$ 
labels LW multiplets. In general, the Lee-Wick mass $M$ can be different for
different scalar fields, in which case the above formula (\ref{RGlw1})
should be generalized in a straightforward way.

\subsection{Effective Potential Approach in the Higher-Derivative Formulation}

We give a nonzero background value $v$ to the complex scalar field 
$\hat\phi$ and write
\be
\hat\phi = \frac{1}{\sqrt{2}}(\hat\varphi + v - i \hat{a})\,,
\ee
and then proceed to derive the (inverse) propagators in that
background. The zeros of such inverse propagators will occur at the
squared masses $M_\alpha^2(v)$. For the scalar field $\hat\varphi$ we
find the secular equation 
\be
\label{secscalar}
P^{-1}_{\hat\varphi}(p) = p^2 - \hat{m}_\varphi^2 - \frac{p^4}{\hat{M}^2}=0\,,
\ee
with $\hat{m}_\varphi^2\equiv\hat{m}^2+3\hat{\lambda}v^2$. The inverse
propagator for the pseudoscalar field $\hat{a}$ 
is similarly obtained with $\hat{m}_\varphi^2\rightarrow
\hat{m}_a^2\equiv\hat{m}^2+\hat{\lambda}v^2$ but, with the gauge-fixing
as in Eq.~(\ref{Lhd}), 
there is also mixing between $\hat{a}$ and $\partial_\mu
\hat{A}^\mu$. The inverse propagator 
for the $\hat{a}$ - $\hat{A}^\mu$ sector is the matrix
\be
\label{secaA}
\left[
\begin{array}{ccc}
\left(p^2-m_A^2 
-\frac{p^4}{M_A^2}\right)g_{\mu\nu}
+\left(-1+\frac{1}{\xi}+\frac{p^2}{M_A^2}+\frac{m_A^2}{\hat{M}^2}
\right)p_\mu p_\nu & &i m_A p_\nu\left(1-\frac{p^2}{\hat{M}^2}\right)\\
&&\\
-i m_A p_\mu\left(1-\frac{p^2}{\hat{M}^2}\right) & & \hat{m}_a^2 - p^2 + 
\frac{p^4}{\hat{M}^2}
\end{array}
\right]\,, 
\ee 
where $m_A(v)\equiv gv$. Equating the determinant of this
matrix to zero we get the secular equation 
\be
(p^4-p^2M_A^2+m_A^2\hat{M}^2)^3
\left[p^6-p^4\hat{M}^2+p^2\hat{m}_a^2(\hat{M}^2+\xi
  m_A^2)-\xi \hat{m}_a^2M_A^2\hat{M}^2\right]=0\,, 
\ee 
for the pole masses in this sector. We see that this equation splits
into two separate equations, of which one  gives pole mass solutions
with multiplicity 3, corresponding to the different polarizations of a
massive gauge boson. Applying to the secular equations
(\ref{secscalar}) and (\ref{secaA}) the prescription
in Eq.~\eqref{recipe1}  we  immediately obtain 
\be {\mathrm{Tr}}[{\cal M}^2]=
(\hat{M}^2)_{\hat\varphi}+3\left(M_A^2\right)_{\hat{A}_\mu}
+(\hat{M}^2)_{\hat{a}}\,, 
\ee 
where the labels indicate
(with some abuse of notation) the origin of each contribution. This
trace is independent of $v$, as it should be to cancel quadratic
divergences in the scalar mass (see discussion in 
Sec.~\ref{sec:LWVeffB}).  We also obtain 
\be
\begin{aligned}
{\mathrm{Tr}}[{\cal M}^4]& = 
 \left(\hat{M}^4-2\hat{M}^2
\hat{m}^2_\varphi\right)_{\hat\varphi}
+3\left(M_A^4-2m_A^2M_A^2\right)_{\hat{A}_\mu}
+\left[\hat{M}^4-2\hat{m}^2_a(\hat{M}^2+\xi m_A^2)\right]_{\hat{a}}\\
&= (v{\text{-indep.~terms}}) -2(3g^2M_A^2+\xi g^2\hat{m}^2 +4\hat\lambda 
\hat{M}^2)v^2 -2\xi\hat\lambda g^2v^4\,.
\end{aligned}
\ee
It follows that 
\begin{align}
16\pi^2(\beta_{\hat{m}^2}+2\gamma_{\hat\phi}\hat{m}^2) &= 
-2(3g^2M_A^2+\xi g^2\hat{m}^2 +4\hat\lambda \hat{M}^2)\,,\\
16\pi^2(\beta_{\hat\lambda}+4\hat\lambda\gamma_{\hat\phi}) &= 
-4\hat\lambda \xi g^2\,,
\end{align}
in perfect agreement with the results  in
Sec.~\ref{sec:diagHD}, Eqs.~(\ref{gammahat})--(\ref{betamhat}). One can 
also check that, in Landau gauge ($\xi=0$) and for $M_A^2=\hat{M}^2=M^2$, 
these equations are in agreement with the general formulas 
(\ref{RGlw1}) and (\ref{RGlw2}).

\subsection{Effective Potential Approach in the Auxiliary-Field Formulation}
\label{betaVAuxF}
In this formulation we give $\phi$ a background value $v$ and write
\be
\phi = \frac{1}{\sqrt{2}}(\varphi + v - i a)\,,
\ee
while
\be
\tilde\phi = \frac{1}{\sqrt{2}}(\tilde\varphi  - i \tilde{a})\,,
\ee
and then proceed to derive the secular equations for the pole masses
$M_\alpha^2(v)$ in the same way as before. 

There is mixing among the CP-even scalars $\varphi$ and
$\tilde\varphi$  and their
inverse propagator is the $2\times 2$ matrix 
\be
\left[
\begin{array}{ccc}
p^2-m_\varphi^2  & & 3\lambda v^2\\
&&\\
3\lambda v^2 & & M^2 - p^2 - 3\lambda v^2 
\end{array}
\right]\,,
\ee
where $m_\varphi^2(h)\equiv m^2 +3\lambda v^2$. Equating the 
determinant
of this matrix to zero, we obtain the secular equation 
\be
\label{secscalaraux}
p^4 - p^2(M^2+m^2) + M^2m^2+3\lambda v^2 (M^2-m^2) = 0\,.
\ee

The fields $A_\mu$, $a$, $\tilde{A}_\mu$ and $\tilde{a}$ get all mixed
in the $v$-background and their inverse propagator is the matrix 
\be
\left[
\begin{array}{ccccccc}
P^{-1}_{\mu\nu}(p) & & i m_A p_\mu & & m_A^2 
g_{\mu\nu}-\frac{1}{\xi}p_\mu p_\nu & & 0\\
& & & & & & \\
- i m_A p_\nu & & m_a^2 - p^2 & & i m_A p_\nu & & - \lambda v^2 \\
& & & & & & \\
m_A^2 g_{\mu\nu}-\frac{1}{\xi}p_\mu p_\nu & & - i m_A p_\mu & & 
\tilde{P}^{-1}_{\mu\nu}(p)& & 0\\
& & & & & & \\
0 & & - \lambda v^2 & & 0 && p^2 -M^2 + \lambda v^2 
\end{array}
\right]\,,
\ee
where 
\begin{align}
P^{-1}_{\mu\nu}(p) & \equiv  
(p^2-m_A^2)g_{\mu\nu}+\left(-1+\frac{1}{\xi}\right)p_\mu p_\nu\,,\\
\tilde{P}^{-1}_{\mu\nu}(p) & \equiv  
(-p^2+M_A^2-m_A^2)g_{\mu\nu}+\left(1+\frac{1}{\xi}\right)p_\mu p_\nu \,,
\end{align}
which leads to the secular equations
\be
\begin{aligned}
\label{secaAaux}
0&= (p^4-p^2M_A^2+m_A^2M_A^2)^3\,,\\
0&= p^6-p^4(M^2+m^2)+p^2\left[m^4+m_a^2(M^2-m^2)+\xi m_a^2 m_A^2\right]\\
&-\xi m_A^2\left[m^4+(M^2-m^2)m_a^2\right]\,.
\end{aligned}
\ee
Applying again to the 
secular equations (\ref{secscalaraux}) and (\ref{secaAaux}) the prescription
in Eq.~\eqref{recipe1} we immediately
obtain 
\be
{\mathrm{Tr}}[{\cal M}^2]= \left(M^2+m^2\right)_{\varphi - 
\tilde\varphi}+3\left(M_A^2\right)_{A_\mu-\tilde{A}_\mu}+\left(M^2+m^2\right)_{a-\tilde{a}}\,,
\ee
where the labels indicate (again with some abuse of notation) the
origin of each contribution. This trace is independent of $v$, as it
should be if the  
quadratic divergences in the scalar mass are to cancel (see discussion in
Sec.~\ref{sec:LWVeffB}). We also obtain  
\begin{align}
{\mathrm{Tr}}[{\cal M}^4] &= \left[M^4+m^4-6\lambda 
v^2(M^2-m^2)\right]_{\varphi - \tilde\varphi}+3\left(M_A^4-2m_A^2 
M_A^2\right)_{A_\mu-\tilde{A}_\mu}\nonumber\\
&\quad+\left[M^4-2\lambda v^2 M^2 
+2m^2\left(m^2+\lambda v^2\right)-2\xi 
m_A^2\left(m^2+\lambda v^2\right)\right]_{a-\tilde{a}}\nonumber\\
&= (v{\text{-indep.}}) -2[3g^2M_A^2+\xi g^2m^2 +4\lambda (M^2-m^2)]v^2 
-2\xi\lambda g^2v^4\,.
\label{tr4aux}
\end{align}

There is now a subtlety when using the scale-independence of the
effective potential due to the fact that, even if the field
$\tilde\phi$ has no background expectation value, its derivative with
the renormalization scale, $d\, \tilde\phi/d\, \log Q$ will have a
nonzero background value that arises from  mixing with the
field $\phi$.  That is, from the tree-level potential
\be
V_0 = m^2 |\phi|^2 - M^2 |\tilde\phi|^2 
+\lambda|\phi-\tilde\phi|^4\,,
\ee
we obtain
\be
\frac{d\, V_0}{d\, \log Q}  = \frac{1}{2}(\beta_{m^2} + 2 
\gamma_{\phi\phi}m^2)v^2 
+\frac{1}{4}\left[\beta_\lambda 
+4\lambda(\gamma_{\phi\phi}-\gamma_{\phi\tilde\phi})\right]v^4\,,
\ee
where $\gamma_{\phi\phi}$ and $\gamma_{\phi\tilde\phi}$ can be read off 
Eq.~(\ref{gammaux}).
Using the previous result for ${\mathrm{Tr}}[{\cal M}^4]$,
Eq.~(\ref{tr4aux}), which determines the scale-dependence of the one-loop
Coleman-Weinberg correction, 
we arrive at 
\begin{align}
16\pi^2(\beta_{m^2}+2\gamma_{\phi\phi}m^2) &= -2\left[3g^2M_A^2+\xi g^2 
m^2 +4\lambda(M^2-m^2)\right]\,,\\
16\pi^2\left[\beta_\lambda+4\lambda(\gamma_{\phi\phi}
-\gamma_{\tilde\phi\phi})\right] &= -4\lambda \xi g^2\,,
\end{align}
in perfect agreement with the results presented in 
Eqs.~(\ref{betaM})-(\ref{betal}).

\section{Some Implications of the UV Behaviour of the LW Standard Model}
\label{sec:disc}

\subsection{Implications at Zero Temperature}
\label{sec:impl0}
We have seen that the LW effective potential
is softer than in standard theories: on the one hand, the bosonic part
of the effective potential, Eq.~(\ref{VblwUV}), does not contain a
$m_\alpha^4\log\Lambda^2$ term while, on the other hand, the fermionic
part, Eq.~(\ref{Vflw}), is finite.  The softer UV behaviour has direct
implications for the RGEs of the LW theory above the threshold
$M$. Using (\ref{RGsm1}) and (\ref{RGsm2}), the RGEs in the SM, using 
Landau gauge, satisfy
\begin{align}
\label{RGsm1exp}
16\pi^2(\beta^{SM}_{\mu^2}+2\gamma^{SM} \mu^2 ) & =  12\lambda \mu^2\ ,
\\
\label{RGsm2exp}
16\pi^2(\beta^{SM}_{\lambda}+4\gamma^{SM} \lambda )& = 
24\lambda^2 - 6 h_t^4 +\frac{3}{4}g^4 + \frac{3}{8}(g^2+{g'}^2)^2\ ,
\end{align}
with the normalization of $\mu^2$ and $\lambda$ as in (\ref{V0}); $g$ 
and $g'$ are the $SU(2)_L$ and $U(1)_Y$ gauge couplings and $h_t$ is the 
top Yukawa coupling. The Higgs anomalous dimension is
\be 
16\pi^2\
\gamma^{\mathrm{SM}}=-3h_t^2 + \frac{3}{4}(3g^2+{g'}^2)\,.  
\ee 
Below the scale $M$ associated with the new LW degrees of freedom these
SM RGEs will still be valid.

Above that scale the full LWSM RGEs should be used. In Landau gauge, we 
can use the same procedure that leads to
(\ref{RGlw1}) and (\ref{RGlw2}) to get
\begin{align}
16\pi^2(\beta_{\hat\mu^2}+2\hat\gamma \hat\mu^2 ) & =  
-\left[12\hat\lambda \hat{M}^2+
\frac{3}{2}(3g^2\hat{M}_A^2+{g'}^2\hat{M}_A^{'2})\right]\ ,
\label{RGlw1exp}
\\
\beta_{\hat\lambda}+4\hat\gamma \hat\lambda  & =  0\ .
\label{RGlw2exp}
\end{align}
The different Lee-Wick masses are the following:
$\hat{M}$ is associated with the Higgs, $\hat{M}_A$ with the $SU(2)_L$ gauge boson
and $\hat{M}'_A$ with the $U(1)_Y$ gauge boson. 
Much as in SUSY theories we see that $\beta_{\hat\lambda}$ is dictated by 
wave-function renormalization only. In particular the SM top-quark
vertex contribution $\sim -h_t^4$ to this beta function
[see Eq.~(\ref{RGsm2exp})] is absent. 

We can easily extend the result for the scalar anomalous dimension in the LW 
Abelian Higgs Model found in a previous section to the Higgs field in the LWSM 
and its non-Abelian gauge structure, simply replacing $g^2 M_A^2$ in 
(\ref{gammahat}) by $\sum_{\gamma A}g_\gamma^2 
T^A_{(\gamma)}T^A_{(\gamma)}M_{A(\gamma)}^2$, where the sum runs over the 
different gauge groups (labeled by $\gamma$) and group generators (labeled by 
$A$), with gauge coupling constant $g_\gamma$ and the $T^A_{(\gamma)}$ are the 
group generator matrices in the representation of the Higgs field. We keep 
explicit the dependence on the different Lee-Wick masses $M_{A(\gamma)}$. In 
contrast with the SM case, this anomalous dimension only gets contributions from 
gauge loops (and not from fermions). In Landau gauge it reads:
\be
16\pi^2\ \hat\gamma=-\frac{3}{2 \hat{M}^2}(3g^2\hat{M}_A^2+{g'}^2\hat{M}^{'2}_A)\,.  
\ee 
In these formulas for the LWSM RGEs we are implicitly adopting the 
higher-derivative formulation. Even if one is interested in a simplified case 
with $\hat{M}=\hat{M}_A=\hat{M}'_A\equiv M$, this condition is not stable under 
RG evolution. The RGEs for the Lee-Wick masses are simple to obtain. Following 
the results of \cite{LWren}, we know that the combinations $g^2 \hat{M}_A^2$ and 
${g'}^2 \hat{M}^{'2}_A$ are scale-invariant in Landau gauge. Therefore, the 
running of the gauge Lee-Wick masses is governed by the evolution of the 
corresponding gauge couplings, which are given explicitly by \cite{LWren}
\be 
\label{betags}
8\pi^2\beta_{g^2}=-2g^4\,,\quad\quad
8\pi^2\beta_{{g'}^2}=\frac{61}{3}{g'}^4\,.  
\ee 
For the RGE of $\hat{M}$ we can generalize the Abelian Higgs case in 
(\ref{betaMhat}) to
\be
\label{betaM2LWSM}
\beta_{\hat{M}^2} = -\frac{3}{16\pi^2}(3g^2 \hat{M}_A^2+{g'}^2\hat{M}^{'2}_A)\,,
\ee
which can be readily integrated.

Focusing on the evolution of the Higgs quartic coupling, we find that 
its scale running in the LWSM above the LW mass is governed by the RGE: 
\be
8\pi^2\beta_{\hat\lambda}= 
3\frac{\hat\lambda}{\hat{M}^2}(3g^2\hat{M}_A^2+{g'}^{2}\hat{M}^{'2}_A)\,.  
\ee 
In leading-log approximation\footnote{In fact, following \cite{LWren}, we 
expect that these beta functions will not receive further contributions beyond 
one loop, with the exception of $\hat\gamma$, which will still be corrected at 
two-loop order.} it is straightforward to integrate this RGE to obtain \be
\label{lloghl}
\hat\lambda(Q>M)=\hat\lambda(M)\left[\frac{M^2}{M^2-
\frac{3}{16\pi^2}(3g^2\hat{M}_A^2+{g'}^{2}\hat{M}^{'2}_A)\log(Q/M)}\right]^2
\,,
\ee 
where $M$ is the common Lee-Wick mass (at the scale $M$).

One consequence of this scale dependence is that $\hat{\lambda}(Q)\geq
\hat{\lambda}(M)$ and that the Higgs effective potential in the LWSM (in
contrast with the SM case) will not develop pathologies at high
scales.
\begin{figure}
\center{
\includegraphics[width=10cm,height=8cm]{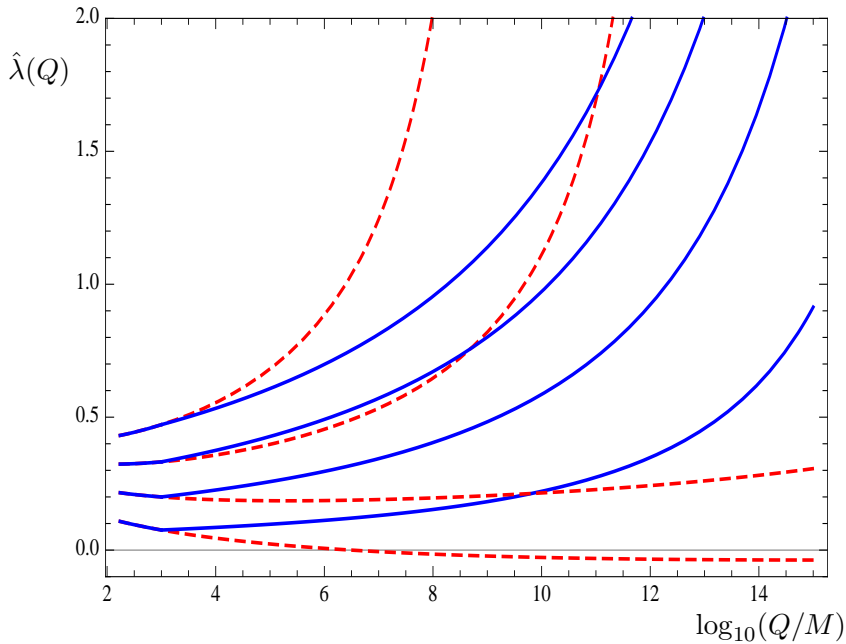}}
\put(-310,200){$\hat\lambda(Q)$}
\put(-50,-10){$\log_{10}(Q/M)$}
\caption{\label{fig:lambdaQ} 
Higgs quartic coupling $\hat\lambda$ running with the renormalization 
scale 
$Q$ in the LW Standard Model (blue solid lines) as compared to the SM (red 
dashed lines) for several values of the Higgs mass. The Lee-Wick mass is 
$M=1$ TeV (note the kink in the RG evolution at that threshold).} 
\end{figure} This is shown in Fig.~\ref{fig:lambdaQ}, which plots the 
running $\hat{\lambda}(Q)$ (for several Higgs mass choices) in the LW 
Standard 
Model (blue solid lines) departing above a Lee-Wick mass $M=1$ TeV from 
the running in the SM (red dashed lines). The plot shows the well known 
fact that in the pure SM, if the Higgs is too light, the running 
$\lambda(Q)$ turns negative at high energies triggering an instability in 
the effective potential. Alternatively, if the Higgs is too heavy, 
$\lambda$ runs into a Landau pole below the Planck scale. For the most 
updated study on this UV fate of the SM and references to the literature, 
see \cite{fate}. In the LW Standard Model, in contrast, the light Higgs 
instability does not take place (provided the LW mass is below the SM 
instability scale) because $\beta_{\hat\lambda}$ is proportional to 
$\hat\lambda$ itself. On the other hand, the heavy Higgs 
non-perturbative regime is 
pushed toward higher masses because $\beta_{\hat\lambda}$ does not grow 
quadratically with $\hat\lambda$ as it does in the SM. In fact the 
explicit solution (\ref{lloghl}) tells us that $\hat\lambda$ hits a Landau 
pole at
\be
\Lambda = M\ 
 {\mathrm{Exp}}\left[\frac{16M^2\pi^2}{3(3g^2\hat{M}_A^2
+{g'}^{2}\hat{M}^{'2}_A)}\right] \ ,
\ee
independently of the Higgs mass value. This means in particular that there 
is no perturbativity bound on the Higgs mass in the LWSM: one could always 
require $\hat{\lambda}(Q\leq M_{Pl})\leq 2\pi$, but the obtained bound 
would not be competitive with the usual unitarity bound, and we do not 
calculate it.\footnote{Lattice 
studies of such bound in similar models, with a higher-derivative kinetic 
term as regulator, exist \cite{Jansen} and show a large increase of the 
bound with respect to the standard case. However, the studied cases use a 
$\phi \partial^6 \phi/M^4$ term, which is higher order than ours, and do 
not include gauge fields, preventing a direct comparison.} Inspection of 
the beta function for $\hat{M}$, (\ref{betaM2LWSM}), also shows that 
$\hat{M}\rightarrow 0$ at the same scale $\Lambda$, which would also be a 
pathological behaviour. At any rate, the numerical value of this cutoff 
scale is higher than the Planck mass and is no cause of concern.

In the previous discussion we have used the coupling $\hat\lambda$, from the 
higher-derivative formulation but similar conclusions follow if we use the 
auxiliary formulation instead. In that formulation, the RGE for the quartic 
coupling $\lambda$ is now
\be
\beta_\lambda=-\frac{48}{16\pi^2}\lambda^2\ ,
\ee
corresponding to a well-behaved, asymptotically-free coupling. In 
agreement with the previous results one cannot obtain lower or upper 
bounds on the Higgs mass on the basis of this running behaviour. 
Nevertheless, the cutoff scale $\Lambda$ reappears in this formulation 
when looking at the running of $M^2+\mu^2$, which goes to zero at that 
scale.

On the other hand, the $U(1)_Y$ gauge coupling $g'$ now runs faster than 
in the SM, see (\ref{betags}), and can become nonperturbative below the 
Planck mass. The Landau pole for this gauge coupling occurs at
\be
\Lambda'\equiv M\ {\mathrm{Exp}}\left[\frac{24\pi^2}{61 
{g'}^{2}(M)}\right]\ .
\ee
For $M=1$ TeV, one gets $\Lambda'\sim 10^{16}$ GeV. This indicates that 
new 
physics beyond the LWSM should appear below $M_{Pl}$. Alternatively, this 
Landau pole could be pushed beyond the Planck mass if the Lee-Wick mass is 
higher, but the required value, of order $M\sim 10^8$ GeV is orders of 
magnitude too high to solve the hierarchy problem.

\subsection{Implications at Finite Temperature}
\label{sec:implT}
As discussed in the introduction, one possible way of probing the acausal
nature of LW theories in search of a macroscopic effect or some 
pathological behaviour is to study them at finite temperature. 
The behavior of a LW gas in thermal equilibrium was studied in
Ref.~\cite{Fornal:2009xc}. It was found there that the contribution to the
free energy $(\Delta\Omega)_{LW}$ of each LW state, that is, of the 
narrow resonances that
would be states of negative metric in the limit that interactions are
switched off, is the negative of the contribution of a normal state of
the same mass:
\begin{equation}
(\Delta\Omega)_{LW}/VT=\begin{cases}
-\int\!\frac{d^3p}{(2\pi)^3}\log\left(1-e^{-E/T}\right),&\text{for 
bosons,}\\
\int\!\frac{d^3p}{(2\pi)^3}\log\left(1+e^{-E/T}\right),&\text{for 
fermions.}
\end{cases}
\label{result}
\end{equation}
Here $E=\sqrt{p^2+M^2}$ and $V$ and $T$ denote volume and
temperature. Consider the energy density $\rho$ at high temperature. For 
each normal scalar degree of freedom (labeled $\alpha$) giving a normal 
contribution with mass $M_{B\alpha1}$ there is a LW
contribution of mass $M_{B\alpha2}$, {\it cf.} Eq.~\eqref{massesb}. 
A high-temperature expansion shows that each bosonic LW multiplet gives a
contribution to the energy density:
\be
\begin{aligned}
\rho_\alpha ^B&
=\left[\frac{\pi^2T^4}{30}-\frac{M_{B\alpha1}^2T^2}{24}+\cdots\right]-
  \left[\frac{\pi^2T^4}{30}-\frac{M_{B\alpha2}^2T^2}{24}+\cdots\right]\\
&=\frac{(M^2-2m_{\alpha}^2)T^2}{24}+\cdots
\end{aligned}
\ee
where we have assumed $m_\alpha<2M$ and used the mass expansions 
of Eq.~(\ref{massesbexp}) in the last step. 
Although the normally leading term $T^4$ is missing, the energy
density is positive and increases with temperature. 

By contrast, the contribution to the energy density of a fermionic LW 
multiplet includes a normal contribution with mass
$M_{F\alpha1}$ and two additional contributions from LW modes of 
masses $M_{F\alpha2,3}$, {\it cf.} Eq.~\eqref{massesf}, with the opposite 
sign. The energy density at high temperature
is dominated by the $T^4$ term and is given by:
\begin{equation}
\begin{aligned}
\rho_\alpha^F&=
\left[\frac{7\pi^2T^4}{240}-\frac{M_{B\alpha1}^2T^2}{48}+\cdots\right]
-\sum_{i=2}^3
\left[\frac{7\pi^2T^4}{240}-\frac{M_{B\alpha i}^2T^2}{48}+\cdots\right]\\
&=-\frac{7\pi^2T^4}{240}
 +\frac{(M^2-m_\alpha^2)T^2}{24}+\cdots
\end{aligned}
\end{equation}
The energy density decreases with temperature and at high enough
temperatures turns negative. This peculiar behavior suggests that, either
interesting phenomena are taking place in the LW fermionic gas at
high temperature or the result (\ref{result}) is not correct, see below. 

We have not computed the effective potential for the scalar field in a
plasma at finite temperature. But there is a well known correspondence
between the zero temperature self-energy diagrams that exhibit
quadratic divergences and the diagrams responsible for a scalar
thermal mass \cite{Comelli:1996vm}. If $\Lambda$ is a straight
momentum cut-off, quadratic divergences in the scalar mass arising
from bosonic excitations, $\Delta m^2=\kappa \Lambda^2/(16\pi^2)$,
translate into a thermal mass correction $\Delta m^2=\kappa
T^2/12$. Similarly, for fermionic excitations $\Delta m^2=-\kappa
\Lambda^2/(16\pi^2)$ translate into $\Delta m^2=\kappa
T^2/24$. Therefore, in models that solve the hierarchy problem by
cancellations of the quadratic divergence in the Higgs mass arising
from intermediate states of the same spin, one expects a corresponding
cancellation in the thermal mass \cite{Espinosa:2004pn}.

The cancellation of quadratically-divergent contributions to the scalar
potential was shown explicitly in Secs.~\ref{sec:LWVeffB} and
~\ref{sec:LWVeffF} for the bosonic and fermionic cases,
respectively. Consider first the bosonic case. The effective
potential, given in Eq.~\eqref{eq:VFromtwoMBs}, 
is the sum of  two same ``normal'' sign contributions. The mass shift
can be obtained by differentiation
\be
\Delta m^2=2\left.\frac{\partial V_1}{\partial v^2}\right|_{0}
\!\!=\sum_\alpha \frac{N_\alpha}{16\pi^2} 
\int_0^{\Lambda^2}\!\!p_E^2\,dp_E^2 
\left.\left[ \frac1{p_E^2+M_{B \alpha 1}^2}\frac{\partial M_{B
      \alpha1}^2}{\partial v^2} 
+  \frac1{p_E^2+M_{B \alpha 2}^2}\frac{\partial M_{B \alpha 
2}^2}{\partial v^2}\right]\right|_{0}\, ,
\ee
where the 0 subscript indicates evaluation at $v=0$.
Since $M_{B \alpha 1}^2+M_{B \alpha 2}^2=M^2$, and $M$ is independent
of the background field, one has
\be
\Delta m^2=\sum_\alpha \frac{N_\alpha}{16\pi^2} 
\int_0^{\Lambda^2}\!\!p_E^2\ 
dp_E^2 \left[ \frac1{p_E^2+M_{B \alpha 1}^2}-\frac1{p_E^2+M^2-M_{B \alpha
      1}^2}\right]\left.\frac{\partial M_{B \alpha 1}^2}{\partial 
v^2}\right|_{0},
\ee
which shows explicitly the cancellation of quadratic
divergences. Rather than performing the angular  momentum
integral that gives Eq.~\eqref{eq:VFromtwoMBs}, one can do first
the integral over the time component of momentum, yielding
\be
V_1 = \sum_\alpha \frac{N_\alpha}{16\pi^3}\int d^3p\; 
(E_{B\alpha1}+E_{B\alpha2})\,,
\ee
where $E_{B\alpha i}=\sqrt{p^2+M_{B \alpha i}^2}$.
The
connection with the finite temperature potential is made, at least in
the normal case, by replacing the energy integral by a sum over
Matsubara modes. Doing this for the LW model, disregarding any
subtleties that may  arise from the LW and CLOP prescriptions, the finite
temperature potential is 
\be
V_1^T = \sum_\alpha \frac{N_\alpha}{16\pi^3}\int d^3p\; 
\left\{\big(E_{B\alpha1}+E_{B\alpha2}\big)
+T\left[\log(1-e^{-E_{B\alpha1}/T})+
\log(1-e^{-E_{B\alpha2}/T})\right]\right\}\,,
\ee
Taking a derivative we obtain the mass shift:
\begin{multline}
\Delta m^2 = \sum_\alpha \frac{N_\alpha}{16\pi^3}\int d^3p\; 
\frac{\partial M_{B \alpha 1}^2}{\partial 
v^2}\left[\left(\frac1{E_{B\alpha1}}-\frac1{E_{B\alpha2}}\right)\right.\\
\left.\left.
+\left(\frac{1}{E_{B\alpha1}}\frac1{e^{E_{B\alpha1}/T}-1}
-\frac{1}{E_{B\alpha2}}\frac1{e^{E_{B\alpha2}/T}-1}
\right)\right]\right|_{0}\,.
\end{multline}
Whilst this expression is not fully justified, it does produce the
expected results, namely  the cut-off independence that
takes place as a cancellation of the $T=0$ terms as well as the
absence of the thermal  $T^2$ mass shift.  But, remarkably, it was
obtained from an effective potential in which the normal and LW modes
enter with normal signs. This is in contrast with the computation of the
free energy in Ref.~\cite{Fornal:2009xc} in which the LW modes appear
with negative sign. However, we have not been able to find any problem 
with the derivation in \cite{Fornal:2009xc} and, at present, we do not 
know which one of these two results, if either, is correct. LW theory is
remarkably intricate and it is possible that missed subtleties have
rendered one or the other calculations, or both,  incorrect. 

The result carries over to the fermionic case. Although there are two
LW modes for one normal mode, the sum rule $\sum_{i=1}^3 M_{F\alpha
  i}^2=2M^2$ produces the cancellations that are associated with the
non-normal signs even though the potential is the sum of normal sign
contributions. Therefore if, contrary to the findings of 
Ref.~\cite{Fornal:2009xc}, LW fields contribute to the thermal free-energy
with normal signs, one would avoid the problem with a negative 
fermionic contribution to the energy density discussed before.

We postpone investigation of the properties of this thermal potential
until a future time when we understand how to better justify the
calculation.

\section*{Acknowledgments}

\noindent 
The authors would like to thank CERN for hospitality and financial 
support. The work of J.R.E was supported in part by the European 
Commission under the European Union through the Marie Curie Research and 
Training Network ``UniverseNet'' (MRTN-CT-2006-035863); by the Spanish 
Consolider-Ingenio 2010 Programme CPAN (CSD2007-00042); and by CICYT, 
Spain, under contracts FPA 2007-60252 and FPA 2010-17747. The work of 
B.G.\ was supported in part by the U.S.\ Department of Energy under 
contract DE-FG03-97ER40546.

\end{document}